\documentclass[11pt]{article}
\usepackage{amsmath}

\newcommand{\N}{N\raise.7ex\hbox{\underline{$\circ $}}$\;$}
\begin{document}

\title{ Maxwell equations in  Riemannian space-time,\\ geometry effect on
material equations in media
}
\author{ V.M. Red'kov$^{2}$,  N.G. Tokarevskaya$^{2}$ ,  E.M.
Bychkouskaya$^{2}$,George J. Spix$^{1}$
%\thanks{E-mail: gjspix@msn.com , redkov@dragon.bas-net.by}
\\
{\small $^1$   Institute of Physics, National  Academy of Sciences
of Belarus  }
\\
{\small $^2$   BSEE Illinois Institute of Technology, USA} }
\date{}
\maketitle

\begin{abstract}

In the paper, the known possibility  to consider the  (vacuum)
Maxwell equations in  a curved space-time as Maxwell equations in
flat space-time (Mandel'stam L.I., Tamm I.E. [1,2])  but taken in
an effective   media the properties of which are determined by
metrical structure of the initial curved model $g_{\alpha\beta}(x)$  is   studied
\begin{eqnarray}
H^{\rho  \sigma }(x)  = \sqrt{-g(x)}\; g^{\rho \alpha } (x) g^{\sigma
\beta}(x) \; \left [ \; \epsilon_{0} \;
 \Delta_{\alpha \beta}^{\;\;\;\;\; \mu\nu} F_{\mu\nu}(x) \; \right ] \; ;
\nonumber
\end{eqnarray}

\noindent $\Delta_{\alpha \beta}^{\;\;\;\;\; ab}$ -- 4-rank tensor;
 metrical  structure of   the
curved space-time  generates effective material equations\hspace{1mm}
for electromagnetic fields:
\begin{eqnarray}
{\bf D}= \epsilon_{0} \;\epsilon(x) \; {\bf E}+ \epsilon_{0}c \;\alpha(x) \; {\bf B} \; ,
\nonumber
\\
{\bf H}= \epsilon_{0}c \;\beta(x) \; {\bf E}+ {1 \over \mu_{0}} \;\mu(x) \; {\bf B} \; ,
\nonumber
\end{eqnarray}

\noindent the form of four symmetrical  tensors $ \epsilon^{ik}(x),
\alpha^{ik}(x),\beta^{ik}(x),\mu^{ik}(x) $ is found explicitly for
general case of  an arbitrary Riemannian space-time geometry
$g_{\alpha \beta} (x)$:
\begin{eqnarray}
\epsilon^{ik}(x)=  \sqrt{-g} \; [ g^{00}(x)
g^{ik} (x) - g^{0i}(x) g^{0k}(x)   ] ,
\qquad \alpha^{ik}(x)= + \sqrt{-g}\;  g^{ij} (x) \;
g^{0l}(x) \; \epsilon_{ljk}
 ,
\nonumber
\\
\beta^{ik} (x) = -  \; \sqrt{-g}  \; g^{0j}(x) \;
\epsilon_{jil} \; g^{lk} (x)
\; , \qquad
\mu^{ik} (x) = \sqrt{-g}  \; {1 \over 2}
\epsilon_{imn} g^{ml} (x) g^{nj} (x)   \epsilon_{ljk}
\; .
\nonumber
\end{eqnarray}

The main  peculiarity  of the geometrical  generating  for
effective electromagnetic  medias characteristics consists in the
following: four tensors $ \epsilon^{ik}(x),
\alpha^{ik}(x),\beta^{ik}(x),\mu^{ik}(x) $ are not independent and
obey some additional constraints between them.Several, the most
simple examples are specified in detail: it is given geometrical
modeling  of  the anisotropic media (magnetic crystals)  and the
geometrical  modeling of a uniform media in moving reference frame
in the background of Minkowsky electrodynamics --  the latter is
realized trough the use of a non-diagonal  metrical tensor
determined by 4-vector  velocity of the  moving uniform media $
g^{am}  =
 [ \; g^{am} + (\epsilon \mu -1) \;  u^{a}u^{m}  \; ] / \sqrt{\mu }\;
$ which gives
\begin{eqnarray}
{\bf D} = + {\epsilon_{0} \over \mu } \;  {\bf E}   +
{\epsilon_{0}  \gamma \over \mu } \;\; {     {\bf E } -
  ({\bf V} {\bf E})  \; {\bf V}   \over 1 - V^{2}}  +
{\epsilon _{0} c  \gamma \over \mu }    \;{  {\bf   V} \times {\bf
B} \over  1 - V^{2}}  \; ,
\nonumber
\\
{\bf H} =   {1  \over  \mu _{0} \mu }\;  {\bf B}     +  {\gamma
\over  \mu _{0} \mu } { {\bf V} \times ({\bf V} \times {\bf B} ) )
\over 1 -V^{2}} + {\epsilon _{0} c  \; \gamma \over \mu  }  \; {
{\bf V} \times {\bf E} \over 1 -V^{2}}
\nonumber
\end{eqnarray}

Also the effective material equations generated by  geometry of space of constant curvature
(Lobachevsky and Riemann models)  are determined.
General problem of  geometrical transforming   arbitrary (linear) material equations, given by
$\epsilon^{(0)}, \alpha^{(0)}, \beta^{(0)}, \mu^{(0)}$,
has been studied -- corresponding  formulas have been produced:
\begin{eqnarray}
{\bf D}  =  \epsilon_{0} \; \left [ \epsilon (x) \;
\epsilon^{(0)}  + \alpha (x)  \beta^{(0)}    \right  ]   \;{\bf  E
} + \epsilon_{0}c \; \left [ \epsilon(x) \alpha^{(0)}  + \alpha(x)
\mu^{(0)}  \right ] \; {\bf B }\; , \nonumber
\\
{\bf H } = \epsilon_{0}  c  \;     \; \left [  \beta (x)
\epsilon^{(0)}  + \mu (x) \beta^{(0)}  \right ] \; {\bf E}   + {1
\over \mu_{0}} \left [ \beta(x) \alpha^{(0)} + \mu(x) \mu^{(0)}
\right ] {\bf B} \; .
\nonumber
\end{eqnarray}

\end{abstract}

\section{ Riemannian geometry and Maxwell theory }

Let us start with the Maxwell equations in Minkowski space: in
vector notation they are [3-6]
\begin{eqnarray}
(I) \qquad \mbox{div}\; {\bf B} = 0 \; , \qquad \mbox{rot}\; {\bf
E} = - {\partial {\bf B} \over  \partial t} \; , \nonumber
\\
(II) \qquad \epsilon \epsilon_{0} \; \mbox{div}\; {\bf E} = \rho
\; , \qquad {1 \over \mu \mu_{0}} \; \mbox{rot}\; {\bf B} = {\bf
J} + \epsilon \epsilon_{0} {\partial  {\bf E}\over \partial t } \;
. \label{1}
\end{eqnarray}

With the use of material equations
\begin{eqnarray}
{\bf H} = { {\bf B} \over  \mu \mu_{0}  }\; , \qquad {\bf D} =
\epsilon \epsilon_{0}\; {\bf E} \; \label{2}
\end{eqnarray}

\noindent  eqs. (\ref{1})  can be written in terms of  four
vectors as follows
\begin{eqnarray}
(I) \qquad \mbox{div}\; c {\bf B} = 0 \; , \qquad \mbox{rot}\;
{\bf E} = - {\partial c {\bf B} \over  \partial x^{0} } \; ,
\nonumber
\\
(II) \qquad  \mbox{div}\; {\bf D} = j^{0} \; , \qquad
 \mbox{rot}\; {{\bf H}\over c} = {\bf j} + {\partial  {\bf D}\over \partial x^{0} }
\label{3}
\end{eqnarray}

\noindent where $ x^{0}= c t\; , j^{a} = (\rho, {\bf J}/c) \; , $
In terms of two electromagnetic tensors:
\begin{eqnarray}
(F^{\alpha \beta} ) = \left | \begin{array}{cccc}
0  & -E^{1}  & -E^{2} & -E^{3} \\
+E^{1}   &  0  &  -cB^{3}  & +  cB^{2}\\
+E^{2} & + cB^{3}  & 0 & - cB^{1} \\
+E^{3} & - cB^{2}& + cB^{1}& 0
\end{array} \right | \; , \;
(H^{\alpha \beta}) = \left | \begin{array}{cccc}
0  & -D^{1}  & -D^{2} & -D^{3} \\
+D^{1}   &  0  &  -H^{3}/c  & +  H^{2}/c \\
+D^{2} & + H^{3}/c  & 0 & - H^{1}/c \\
+D^{3} & - H^{2}/c & +H^{1}/c & 0
\end{array} \right |
%\label{4}
\nonumber
\end{eqnarray}

\noindent eqs. (\ref{3}) take the form
\begin{eqnarray}
(I) \qquad
\partial _{a} F_{bc}  +  \partial _{b} F_{ca}  + \partial _{c} F_{ab}=
0\; , \qquad (II) \qquad \qquad
\partial_{b}H^{ba} = j^{a}\; .
 \label{5}
 \end{eqnarray}

In vacuum case, the material equations\footnote{Note the notation
$ E^{i}= - E_{i} , \; D^{i}= - D_{i} , \; B^{i} = + B_{i}  ,\;
 H^{i} = + H_{i} $}
$$
{\bf D} = \epsilon_{0} {\bf E}=(D^{i}) \; , \qquad
 {\bf H} = {1 \over \mu_{0}}\;  {\bf B} = (H^{i}),
$$

\noindent will look in tensor form as follows:
$$
H^{ab}(x) = \epsilon_{0}\; F^{a b} (x) \; .
$$

\noindent
The situation is  quite different in non-vacuum case. For
instance, the material equations for a uniform media
$$
{\bf D} = \epsilon_{0} \epsilon  {\bf E}=(D^{i}) \; , \qquad
 {\bf H} = {1 \over \mu_{0}\mu }\;  {\bf B} = (H^{i}) ,
$$

\noindent these relationships can be written in short form with
the help of subsidiary $4\times 4$ - matrix
\begin{eqnarray}
\eta^{am} = \sqrt{\epsilon} \; \left | \begin{array}{cccc}
1/k & 0 & 0 & 0 \\
0 & -k &  0 & 0 \\
0 & 0 & -k & 0 \\
0 & 0 & 0& -k
\end{array} \right | \; , \qquad k ={1 \over \sqrt{\epsilon \mu}}
, \qquad H^{a b} = \epsilon_{0}  \; \eta^{am} \eta^{bn} \; F_{mn}
\label{d}
\end{eqnarray}

When extending Maxwell theory to the case of space-time with
non-Euclidean geometry, which can describe gravity according to
General Relativity [6], one must change previous equations  to a
more general form [6] (for simplicity, let us start with the most
simple case of vacuum Maxwell equations):
\begin{eqnarray}
(I) \qquad \nabla_{\alpha} F_{\beta \gamma} + \nabla_{\beta}
F_{\gamma \alpha} + \nabla_{\gamma} F_{\alpha \beta } = 0 \; ,
\nonumber
\\
(II) \qquad \nabla_{\beta}H^{ \beta \alpha} = j^{\alpha} \;  ,
\qquad H_{\alpha \beta} = \epsilon _{0} \; F_{\alpha \beta} \; .
\label{6}
\end{eqnarray}

\section{ Maxwell equations in Riemannian space-time and a media}

Let us discuss in detail the known possibility [1-2] to consider
the  (vacuum) Maxwell equations in  a curved space-time as Maxwell
equations in flat space-time but taken in an effective   media the
properties of which are determined by metrical structure of the
initial curved model $g_{\alpha\beta}(x)$. Let us restrict
ourselves to the case of curved space-time models which are
parameterized by the same quasi-Cartesian coordinate system
$x^{a}$.

Vacuum Maxwell equations in a Riemannian space-time, parameterized
by the same quasi-Cartesian coordinates (to distinguish formulas
referring to a flat and curved models let us use  small letters to
designates electromagnetic tensors in curved model, $f_{ab}$ and
$h^{ab}$ )
\begin{eqnarray}
(I) \qquad \partial_{a} f_{bc} +
\partial_{b} f_{c a} +
\partial_{c} f_{ab} = 0 \; , \qquad
(II) \qquad
  {1 \over \sqrt{-g}} \; \partial _{b}
\;  \sqrt{-g} \; f^{ b a} = {1 \over \epsilon_{0}}  j^{a} \; .
\label{42}
\end{eqnarray}

\noindent One can immediately see that introducing new (formal)
variables (there  exists one special case; namely, if $g(x)$ does
not depend on coordinates in fact  then the factor $\sqrt{-g}$ can
be omitted from the formulas  and below)
\begin{eqnarray}
F_{ab} = f_{ab} , \qquad H^{ba} =  \epsilon_{0} \;\sqrt{-g} \;
g^{am}(x) g^{bn}(x)\; f_{mn}(x)  , \qquad
   \sqrt{-g} \; j^{a} \longrightarrow  j^{a}
\label{43}
\end{eqnarray}

\noindent equations   (\ref{42}) in the curved space can be
re-written as Maxwell equations of the type (\ref{41a}) in flat
space but in a media:
\begin{eqnarray}
(I) \qquad \partial_{a} F_{bc} +
\partial_{b} F_{ca} +
\partial_{c} F_{ab } = 0
\qquad (II) \qquad
 \partial _{b}
\;   F^{ ba} =  {1 \over \epsilon_{0}} j^{a} \; . \label{44}
\end{eqnarray}

\noindent At this, relations playing the role of material
equations are determined by metrical structure:
\begin{eqnarray}
H^{\beta \alpha}(x)  = \epsilon_{0}\; [\; \sqrt{-g(x)}\;
g^{\alpha \rho} (x) g^{\beta \sigma}(x)\; ]
 \;  F_{\rho \sigma} (x) \; ;
\label{45}
\end{eqnarray}

\noindent if $g_{\alpha\beta} $  does not depend upon coordinates,
then the factor $\sqrt{-g(x)}$ can be omitted    --- see
(\ref{43}).

\section{ Metrical tensor  $g_{\alpha \beta}(x)$ and
material equations}

In this section let us consider the material equations
 for electromagnetic
fields which are generated by metrical structure of the curved
space-time model.
 Consider the case of arbitrary metrical tensor
\begin{eqnarray}
g_{\alpha \beta}(x) = \left | \begin{array}{cccc}
g_{00} & g_{01} & g_{02} & g_{03} \\
g_{01} & g_{11} & g_{12} & g_{13} \\
g_{02} & g_{12} & g_{22} & g_{23}  \\
g_{03} & g_{13} & g_{23} & g_{33}
\end{array} \right | .
\label{53}
\end{eqnarray}

\noindent We are to obtain a 3-dimensional form of relation
(\ref{45}). Their general structure should be as
follows\footnote{For discussion of different types of
electromagnetic medias see in
 [7-12]}:
\begin{eqnarray}
D^{i}= \epsilon_{0}\; \epsilon^{ik}(x) \; E_{k}+
\epsilon_{0}c\;\alpha^{ik}(x) \; B_{k} \; , \nonumber
\\
H^{i}= \epsilon_{0}c \;\beta^{ik}(x) \; E_{k}+ {1 \over  \mu_{0}}
\; \mu^{ik}(x) \; B_{k} \; . \label{54}
\end{eqnarray}

\noindent Four dimensionless $(3\times3)$-matrices
$\epsilon^{ik}(x),\;\alpha^{ik}(x),\; \beta^{ik}(x),\;\mu^{ik}(x)$
should not be independent because they are bilinear functions of
10 independent components of the symmetrical tensor
$g_{\alpha\beta}(x)$. After simple calculation, one produces
expressions for four tensors:
\begin{eqnarray}
\epsilon^{ik}(x)= \sqrt{-g} \; (g^{00}(x) \; g^{ik} (x) -
g^{0i}(x) \; g^{0k}(x) ) \; , \nonumber
\\
\mu^{ik} (x) =    {1 \over 2} \;  \sqrt{-g} \;  \epsilon_{imn} \;
g^{ml} (x) g^{nj} (x)  \; \epsilon_{ljk}  \; , \nonumber
\\
\alpha^{ik}(x)= + \sqrt{-g} \;   g^{ij} (x) \; g^{0l}(x) \;
\epsilon_{ljk} \; , \nonumber
\\
\beta^{ik} (x) =  -  \sqrt{-g} \;  g^{0j}(x) \;  \epsilon_{jil} \;
g^{lk} (x)  \; . \label{57}
\end{eqnarray}

\noindent  The above form the tensors obey special symmetry
conditions:
\begin{eqnarray}
\epsilon^{ik}(x) = + \epsilon^{ki}(x) \; , \qquad \mu^{ik}(x) = +
\mu^{ki}(x) \; , \qquad \beta^{ki}(x) =  \alpha^{ik} \; ;
\label{58a}
\end{eqnarray}

\noindent which mean that the $(6\times 6)$-matrix defining
material equations
\begin{eqnarray}
\left | \begin{array}{c} D^{i}(x) \\ H^{i} (x) \end{array} \right
| = \left | \begin{array}{cc} \epsilon_{0} \; \epsilon^{ik} (x) &
\epsilon_{0}c \; \alpha^{ik} (x) \\ [2mm] \epsilon_{0}c  \;
\beta^{ik} (x) &  \mu^{-1}_{0}\; \mu^{ik} (x)
\end{array} \right |
\left | \begin{array}{c} E_{k}(x) \\ B_{k} (x) \end{array} \right
| \; \label{58b}
\end{eqnarray}

\noindent  is  a symmetrical matrix. Four (material) tensor in the
above formulas are defined by
\begin{eqnarray}
[\; \epsilon^{ik}(x)\;]  =   \sqrt{-g}\; g^{00}\; \left |
\begin{array}{ccc}
g^{11} & g^{12} & g^{13} \\
g^{21} & g^{22} & g^{23} \\
g^{31} & g^{32} & g^{33}
\end{array} \right |  -    \sqrt{-g}\; \left | \begin{array}{ccc}
g^{1}\;g^{1}  &  g^{1}\; g^{2} &  g^{1}\; g^{3} \\
g^{2}\;g^{1} &  g^{2} \;g^{2} &  g^{2}\;g^{3} \\
g^{3} \;g^{1} &  g^{3}\; g^{2} &  g^{3} \;g^{3}
\end{array} \right |  \; ,
\nonumber
\\
\mu^{ik} (x) = ( \sqrt{-g}\;  \left | \begin{array}{ccc} (g^{22}
g^{33} - g^{23} g^{32})  &  (g^{31} g^{23} -  g^{21} g^{33}) &
 (g^{21} g^{32} - g^{22} g^{31})\\
(g^{32} g^{13} -g^{33}g^{12})  & (g^{33}g^{11}- g^{31}g^{13}) &
(g^{31}g^{12} - g^{32}g^{11}) \\
(g^{12} g^{23} - g^{13}g^{22} ) & (g^{13}g^{21} - g^{11}g^{23} ) &
(g^{11}g^{22} - g^{12} g^{21} )
\end{array} \right |\; ,
\nonumber
\\
\alpha^{ik} (x)  = \sqrt{-g}\; \left | \begin{array}{ccc}
(- g^{12} g^{3} +g^{13} g^{2}) & ( g^{11}g^{3} - g^{13} g^{1} ) & (-g^{11}g^{2} + g^{12} g^{1} )\\
(-g^{22}g^{3} + g^{23} g^{2}) & (g^{21}g^{3} - g^{23}g^{1} ) & ( -g^{21}g^{2} + g^{22}g^{1} )\\
(-g^{32}g^{3} + g^{33} g^{2}) &   ( g^{31} g^{3}- g^{33} g^{1} ) &
( -g^{31} g^{2} + g^{32} g^{1} )
\end{array}  \right | ,
\nonumber
\\
\beta^{ik} (x)   =  \sqrt{-g}\;   \left | \begin{array}{ccc}
(- g^{12} g^{3} +g^{13} g^{2}) & (-g^{22}g^{3} + g^{23} g^{2})  & (-g^{32}g^{3} + g^{33} g^{2}) \\
( g^{11}g^{3} - g^{13} g^{1} ) & (g^{21}g^{3} - g^{23}g^{1} ) & ( g^{31} g^{3}- g^{33} g^{1} ) \\
(-g^{11}g^{2} + g^{12} g^{1} ) &   ( -g^{21}g^{2} + g^{22}g^{1} )
& ( -g^{31} g^{2} + g^{32} g^{1} )
\end{array}  \right |  \; .
\label{15.5b}
\end{eqnarray}

\section{ Geometrical modeling of the uniform media}

Let us consider  one special form of the  metrical tensor:
\begin{eqnarray}
g_{\alpha \beta} (x) = \left | \begin{array}{cccc}
a^{2} & 0  & 0 & 0 \\
0 & -b^{2} & 0 & 0\\
0 & 0 & -b^{2} & 0 \\
0 & 0 & 0 & -b^{2}
\end{array} \right | \; ,
\label{91}
\end{eqnarray}

\noindent where $a^{2}$ and $b^{2}$ are  arbitrary (positive)
numerical parameters. This is a special case mentioned in
connection with eq.  (\ref{43}):  if $g(x)$ does not depend on
coordinates in fact  then the factor $\sqrt{-g}$ can be omitted
from the formulas. Acting so we get the material equations
generated by that geometry
\begin{eqnarray}
(\epsilon^{ik}) =   { 1  \over a^{2}b^{2} } \left |
\begin{array}{ccc}
-1 & 0 & 0 \\
0 & -1 & 0 \\
0 & 0 & -1
\end{array} \right |\; ,\qquad
(\mu^{ik} ) =  { 1  \over b ^{4}} \; \left | \begin{array}{ccc}
1 & 0 & 0 \\
0 & 1 & 0 \\
0 & 0 & 1
\end{array} \right | \; ,
\label{92}
\end{eqnarray}

\noindent or differently
\begin{eqnarray}
D^{i} = -{\epsilon_{0}  \over a^{2}b^{2}} \; E_{i}  \;, \qquad
H^{i} = {1 \over  \mu_{0} b^{4}} \; B_{i} \; , \label{93}
\end{eqnarray}

\noindent from which it follows
\begin{eqnarray}
b^{2} = \sqrt{\mu }  \; , \qquad a^{2} = {1\over \epsilon }
 {1 \over \sqrt{  \mu}}\; .
\label{97c}
\end{eqnarray}

\noindent Corresponding metrical tensor (\ref{91})  is
\begin{eqnarray}
g_{\alpha \beta} (x) =
 {1 \over \sqrt{\epsilon  }}
\left | \begin{array}{cccc}
1 /\sqrt{\epsilon  \mu  }  & 0  & 0 & 0 \\
0 & -\sqrt{\epsilon  \mu }  & 0 & 0\\
0 & 0 & - \sqrt{\epsilon  \mu} & 0 \\
0 & 0 & 0 & -\sqrt{\epsilon  \mu }
\end{array} \right |
\; .
%, \qquad \sqrt{-g} = a \; b^{3}= {\mu  \over \epsilon } \; .
\label{98a}
\end{eqnarray}

\section{  Geometrical modeling of an anisotropic media}

Let us extend the previous analysis and consider another  metrical
tensor:
\begin{eqnarray}
g_{\alpha \beta}  = \left | \begin{array}{cccc}
a^{2} & 0  & 0 & 0 \\
0 & -b_{1}^{2} & 0 & 0\\
0 & 0 & -b_{2}^{2} & 0 \\
0 & 0 & 0 & -b_{3}^{2}
\end{array} \right | \; ,
\label{107}
\end{eqnarray}

\noindent where $a^{2}, b_{1}^{2}, b_{2}^{2},b_{3}^{2},$ are
arbitrary  numerical parameters. The material equations generated
by that geometry are
\begin{eqnarray}
D^{i}= \epsilon_{0} \epsilon^{ik} \; E_{k}  \; , \qquad
(\epsilon^{ik}) =    a^{-2} \left | \begin{array}{ccc}
-b_{1}^{-2} & 0 & 0 \\
0 & -b_{2}^{-2} & 0 \\
0 & 0 & -b_{3}^{-2}
\end{array} \right |\;
, \label{108}
\\
H^{i}= \mu_{0}^{-1} \mu^{ik} \; B_{k} \; , \qquad (\mu^{ik} ) =
\left |
\begin{array}{ccc}
b_{2}^{-2}b_{3}^{-2} & 0 & 0 \\
0 & b_{3}^{-2}b_{1}^{-2} & 0 \\
0 & 0 & b_{1}^{-2}b_{2}^{-2}
\end{array} \right | \; ,
\nonumber
\end{eqnarray}

\noindent or differently
\begin{eqnarray}
D^{1} = -{\epsilon_{0}  \over a^{2}b_{1}^{2}} \; E_{1}  \;, \qquad
D^{2} = -{\epsilon_{0} \over a^{2}b_{2}^{2}} \; E_{2}  \;, \qquad
D^{3} = -{\epsilon_{0} \over a^{2}b_{3}^{2}} \; E_{3}  \;,
\nonumber
\\
H^{1} = {1 \over \mu_{0}\; b_{2}^{2}b_{3}^{2}}\; B_{1} \; ,\qquad
H^{2} = {1 \over \mu_{0} \; b_{3}^{2}b_{1}^{2}}\; B_{2} \; ,\qquad
H^{3} = {1 \over \mu_{0}\;  b_{1}^{2}b_{2}^{2}}\; B_{3} \; .
%\label{109}
\nonumber
\end{eqnarray}

These material equations should be  compared with
\begin{eqnarray}
 D^{1}  = - \epsilon_{0} \epsilon _{1} \;  E_{1} \; , \qquad
  D^{2}  = - \epsilon_{0} \epsilon _{2} \;  E_{2} \; , \qquad
 D^{3}  = -  \epsilon_{0} \epsilon _{3} \;  E_{3} \; ,
\nonumber
\\
 H^{1}  = {1 \over \mu_{0}  \mu_{1}} \; B_{1} \;  , \qquad
 H^{2}  = {1 \over  \mu_{0} \mu_{2}} \; B_{2} \;  , \qquad
H^{3}  = {1 \over   \mu_{0} \mu_{3}} \; B_{3} \;  , \nonumber
\end{eqnarray}

\noindent from which it follows
\begin{eqnarray}
\epsilon_{1} = {1  \over a^{2} b^{2}_{1}} \;,\qquad \epsilon_{2} =
{1 \over a^{2} b^{2}_{2}} \;,\qquad \epsilon_{3} = {1 \over a^{2}
b^{2}_{3}} \;,
%\label{110a}
\nonumber
\\
\mu_{1} =   b_{2}^{2}\; b_{3}^{2} \; , \qquad \mu_{2} =
b_{3}^{2}\; b_{1}^{2} \; , \qquad \mu_{3} =  b_{1}^{2}\; b_{2}^{2}
\; . \label{110b}
\end{eqnarray}

\noindent One can readily obtain
\begin{eqnarray}
{\mu_{1} \over \epsilon_{1}} = {\mu_{2}\over \epsilon_{2}} =
{\mu_{3}\over \epsilon_{3}} =
 (a^{2}\;b^{2}_{1}\;b^{2}_{2}\;b^{2}_{3})
= -g  \; , \nonumber
\\
-g  =\sqrt{
 {\mu_{1}^{2} + \mu_{2}^{2} +\mu_{3}^{2}\over
  \epsilon_{1}^{2} + \epsilon_{2}^{2} +\epsilon_{2}^{2}  } } \; ,
\qquad {\mu_{i} \over \sqrt{\mu_{1}^{2} + \mu_{2}^{2}
+\mu_{3}^{2}} } = { \epsilon_{i} \over \sqrt{\epsilon_{1}^{2} +
\epsilon_{2}^{2} +\epsilon_{2}^{2} } } \; . \label{111}
\end{eqnarray}

\noindent
The latter means that one may use four independent
parameters, $\epsilon, \mu, n_{i}$:

\begin{eqnarray}
\epsilon_{i} = \epsilon \; n_{i}\; , \qquad \mu_{i}= \mu \;n_{i},
\qquad {\bf n} ^{2} = 1 \label{112}
\end{eqnarray}

\noindent One can readily express $b_{i}^{2}$ in terms of
$\mu_{i}$:
\begin{eqnarray}
\mu_{2} \mu_{3} =   b_{1}^{4} \;(b_{2}^{2}\; b_{3}^{2}) =
b_{1}^{4} \; \mu_{1} \qquad \Longrightarrow \qquad b_{1}^{2} =
\sqrt{{\mu_{2} \mu_{3} \over \mu_{1}}}=
 \sqrt{\mu} \;  \sqrt{{n_{2} n_{3} \over n_{1}}} \; ,
\nonumber
\\
\mu_{3} \mu_{1} =  b_{2}^{4} \;(b_{3}^{2}\; b_{1}^{2}) =
b_{2}^{4} \; \mu_{2} \qquad \Longrightarrow \qquad b_{2}^{2} =
\sqrt{{\mu_{3} \mu_{1} \over  \mu_{2}}} = \sqrt{\mu}
\;\sqrt{{n_{3} n_{1} \over n_{2}}} \; , \nonumber
\\
\mu_{1} \mu_{2} = \mu_{0}^{2}\; b_{3}^{4} \;(b_{1}^{2}\;
b_{2}^{2}) =\mu_{0} \;  b_{3}^{4} \; \mu_{3} \qquad
\Longrightarrow \qquad b_{3}^{2} = \sqrt{{\mu_{1} \mu_{2} \over
\mu_{3}}} = \sqrt{\mu} \;\sqrt{{n_{1} n_{2} \over n_{3}}} \; .
\nonumber
\\
\label{113a}
\end{eqnarray}

\noindent In turn, from $ a^{2} \; b_{1}^{2}\; b_{2}^{2} \;
b_{3}^{2} =  \mu / \epsilon$ it follows
\begin{eqnarray}
a^{2} = {\mu \over \epsilon} \; {1 \over
b_{1}^{2}b_{2}^{2}b_{3}^{2}}= {1 \over \epsilon  \sqrt{\mu}} \; {1
\over \sqrt{n_{1}n_{2}n_{3}}} \label{113b}
\end{eqnarray}

The formula (\ref{113a})-(\ref{113b})  provide us with
(anisotropic) extension
\begin{eqnarray}
g_{ab}(x) = {1 \over \sqrt{\epsilon}} \left | \begin{array}{cccc}
{1 \over   \sqrt{ \epsilon \mu}} \; {1 \over \sqrt{n_{1}n_{2}n_{3}}}  & 0 & 0 & 0 \\
0 &  -\sqrt{\epsilon \mu} \;  \sqrt{{n_{2} n_{3} \over n_{1}}}
 & 0 & 0 \\
0 & 0 & -\sqrt{\epsilon\mu} \;\sqrt{{n_{3} n_{1} \over n_{2}}} & 0 \\
0 & 0 & 0 & -\sqrt{\epsilon\mu} \;\sqrt{{n_{1} n_{2} \over n_{3}}}
\end{array} \right |
\label{114}
\end{eqnarray}

\noindent of the  previous (isotropic) metrical tensor.

\section{  The moving media and anisotropy }

One other, more involved,  example of effective  anisotropic media
is provided by the material equations for uniform media for a
moving observer (more details see in [14-16]):
\begin{eqnarray}
\Delta^{abmn} =  {\epsilon_{0} \over \mu }  \;
 [ \; g^{am} + (\epsilon \mu -1) \;  u^{a}u^{m}  \; ] \;
 [ \; g^{bn} + (\epsilon \mu -1)\;  u^{b}u^{n}  \;  ] \; , \qquad
H^{ab} (x) =  \Delta^{abmn} \; F_{mn}  \; . \nonumber
\\
\label{10.1}
\end{eqnarray}

Corresponding four 3-dimensional tensors are
\begin{eqnarray}
\epsilon^{ik}= {1 \over  \mu } \left | \begin{array}{ccc} (-1 +
\gamma  u^{1}u^{1} - \gamma  u^{0}u^{0} ) & \gamma  u^{1}u^{2}
 & \gamma  u^{1}u^{3} \\
\gamma  u^{1}u^{2} &
(-1 + \gamma  u^{2}u^{2} - \gamma  u^{0}u^{0} ) & \gamma  u^{2}u^{3}\\
\gamma  u^{3}u^{1} & \gamma  u^{3}u^{2} & (-1 + \gamma  u^{3}u^{3}
- \gamma  u^{0}u^{0} )
\end{array} \right |
, \nonumber
\\[3mm]
\mu^{ik}= {1 \over  \mu } \left | \begin{array}{ccc} (1 - \gamma
u^{2}u^{2} - \gamma  u^{3}u^{3} ) & \gamma  u^{1}u^{2}
 & \gamma  u^{1}u^{3} \\
\gamma  u^{1}u^{2} &
(1 - \gamma  u^{3}u^{3} - \gamma  u^{1}u^{1} ) & \gamma  u^{2}u^{3}\\
\gamma  u^{3}u^{1} & \gamma  u^{3}u^{2} & (1 - \gamma  u^{1}u^{1}
- \gamma  u^{2}u^{2} )
\end{array} \right |,
\nonumber
\\[3mm]
\alpha^{ik}  = { 1  \over \mu } \left | \begin{array}{ccc}
0                        & -  \gamma  u^{0} u^{3}  & +  \gamma  u^{0} u^{2} \\
+  \gamma  u^{0} u^{3} &         0                 & -  \gamma  u^{0} u^{1} \\
- \gamma  u^{0} u^{2}  & + \gamma  u^{0} u^{1}   &              0
\end{array} \right |  , \qquad
\beta^{ik}  = { 1  \over \mu } \left | \begin{array}{ccc}
0                        & +  \gamma  u^{0} u^{3}  & -  \gamma  u^{0} u^{2} \\
-  \gamma  u^{0} u^{3} &         0                 & +  \gamma  u^{0} u^{1} \\
+ \gamma  u^{0} u^{2}  & - \gamma  u^{0} u^{1}   &              0
\end{array} \right | \; .
\nonumber
\\
\end{eqnarray}

Let us deduce 3-dimensional vector form of these relations. For
the vector   $D^{i}$ we have
\begin{eqnarray}
 D^{1} =
{\epsilon_{0} \over \mu } \;  [ \; (-1 + \gamma  u^{1} u^{1} -
\gamma  u^{0} u^{0} ) E_{1} + \gamma u^{1} u^{2} E_{2} + \gamma
u^{1} u^{3} E_{3} ] + {\epsilon _{0} c \over \mu } (- \gamma u^{0}
u^{3} B_{2} + \gamma u^{0} u^{2} B_{3}) \nonumber
\\
D^{2} = {\epsilon_{0} \over \mu } \;  [\;   + \gamma u^{1} u^{2}
E_{1}  + (-1 + \gamma  u^{2} u^{2} - \gamma  u^{0} u^{0} ) E_{2} +
\gamma u^{2} u^{3} E_{3} ] + {\epsilon _{0} c \over \mu } ( \gamma
u^{0}  u^{3} B_{1} - \gamma u^{0} u^{1} B_{3}) \nonumber
\\
D^{1} = {\epsilon_{0} \over \mu } \;  [  + \gamma u^{1} u^{3}
E_{1} +  \gamma u^{2} u^{3} E_{2} + (-1 + \gamma  u^{3} u^{3} -
\gamma  u^{0} u^{0} ) E_{3}  ] + {\epsilon _{0} c \over \mu } (-
\gamma u^{0}  u^{2} B_{1} + \gamma u^{0} u^{1} B_{2}) \nonumber
\end{eqnarray}

\noindent and further
\begin{eqnarray}
D^{1} = - {\epsilon_{0} \over \mu } \;  E_{1}   +  {\epsilon_{0}
\gamma \over \mu } \;  [ - u^{0} u^{0}  E_{1} +
  (u^{1} E_{1}  +   u^{2} E_{2} +    u^{3} E_{3})  \; u^{1}  \; ] +
{\epsilon _{0} c  \gamma \over \mu }    \; u^{0}  \; (  u^{2}
B_{3} -   u^{3} B_{2}) \; , \nonumber
\\
D^{2} = - {\epsilon_{0} \over \mu } \;  E_{2}   +  {\epsilon_{0}
\gamma \over \mu } \;  [ - u^{0} u^{0}  E_{2} +
  (u^{1} E_{1}  +   u^{2} E_{2} +    u^{3} E_{3}) \;  u^{2}  \; ] +
{\epsilon _{0} c  \gamma \over \mu }   \;  u^{0}  \; (  u^{3}
B_{1} -   u^{1} B_{3}) \; , \nonumber
\\
D^{3} = - {\epsilon_{0} \over \mu } \;  E_{3}   +  {\epsilon_{0}
\gamma \over \mu } \;  [ - u^{0} u^{0}  E_{3} +
  (u^{1} E_{1}  +   u^{2} E_{2} +    u^{3} E_{3}) \;  u^{3}  \; ] +
{\epsilon _{0} c  \gamma \over \mu } \;    u^{0}  \; (  u^{1}
B_{2} -   u^{2} B_{1}) \; , \nonumber
\end{eqnarray}

\noindent With the use of  notation
$$
u^{0} = {1 \over \sqrt{1 - V^{2}}  } , \qquad  u^{i} = {V^{i}
\over \sqrt{1 - V^{2}}  }
$$

\noindent  previous relations look as follows
\begin{eqnarray}
D^{1} = - {\epsilon_{0} \over \mu } \;  E_{1}   +  {\epsilon_{0}
\gamma \over \mu } \; { [ -   E_{1} +
  (V^{1} E_{1}  +   V^{2} E_{2} +    V^{3} E_{3})  \; V^{1}  \; ] \over 1 - V^{2}}  +
{\epsilon _{0} c  \gamma \over \mu }    \;{ (  V^{2} B_{3} - V^{3}
B_{2}) \over  1 - V^{2}}  \; ,
\nonumber
\\
D^{2} = - {\epsilon_{0} \over \mu } \;  E_{2}   +  {\epsilon_{0}
\gamma \over \mu } \; { [ -   E_{1} +
  (V^{1} E_{1}  +   V^{2} E_{2} +    V^{3} E_{3})  \; V^{2}  \; ] \over 1 - V^{2}}  +
{\epsilon _{0} c  \gamma \over \mu }    \;{ (  V^{3} B_{1} - V^{1}
B_{3}) \over  1 - V^{2}}  \; ,
\nonumber
\\
D^{3} = - {\epsilon_{0} \over \mu } \;  E_{3}   +  {\epsilon_{0}
\gamma \over \mu } \; { [ -   E_{3} +
  (V^{1} E_{1}  +   V^{2} E_{2} +    V^{3} E_{3})  \; V^{3}  \; ] \over 1 - V^{2}}  +
{\epsilon _{0} c  \gamma \over \mu }    \;{ (  V^{1} B_{2} - V^{2}
B_{1}) \over  1 - V^{2}}  \; ,
\nonumber
 \end{eqnarray}

\noindent or in  vector form they are
\begin{eqnarray}
{\bf D} = + {\epsilon_{0} \over \mu } \;  {\bf E}   +
{\epsilon_{0}  \gamma \over \mu } \;\; {     {\bf E } -
  ({\bf V} {\bf E})  \; {\bf V}   \over 1 - V^{2}}  +
{\epsilon _{0} c  \gamma \over \mu }    \;{  {\bf   V} \times {\bf
B} \over  1 - V^{2}}  \; , \label{130a}
\end{eqnarray}

Now analogously we should consider three relations for
 $H^{i}$:
\begin{eqnarray}
H_{1} = {1  \over  \mu _{0} \mu } \;  [ \; (1 - \gamma  u^{2}
u^{2} - \gamma  u^{3} u^{3} )\;  B_{1}  + \gamma u^{1} u^{2} B_{2}
+  \gamma u^{1} u^{3} B_{3} \; ] + {\epsilon _{0} c \over \mu } (
\gamma u^{0}  u^{3} E_{2} - \gamma u^{0} u^{2} E_{3})
\nonumber
\\
H_{2} = {1  \over  \mu _{0} \mu } \;  [ \; \gamma u^{1} u^{2}
B_{1} +  (1 - \gamma  u^{3} u^{3} - \gamma  u^{1} u^{1} )\;  B_{2}
+  \gamma u^{2} u^{3} B_{3} \; ] + {\epsilon _{0} c \over \mu } (
\gamma u^{0}  u^{3} E_{2} - \gamma u^{0} u^{2} E_{3})
\nonumber
\\
H_{3} = {1  \over  \mu _{0} \mu } \;  [ \; \gamma u^{3} u^{1}
B_{1} +
  \gamma u^{3} u^{2} B_{3}  + (1 - \gamma  u^{1} u^{1} - \gamma  u^{2} u^{2} )\;  B_{3} \; ] +
{\epsilon _{0} c \over \mu } ( \gamma u^{0}  u^{2} E_{1} - \gamma
u^{0} u^{1} E_{2})
\nonumber
\end{eqnarray}

\noindent  or further
\begin{eqnarray}
H_{1} =   {1  \over  \mu _{0} \mu }\;  B_{1}     +  {\gamma \over
\mu _{0} \mu } (-  u^{2} u^{2} \;  B_{1}  - u^{3} u^{3} \; B_{1}
+  u^{1} u^{2} B_{2} +  u^{1} u^{3} B_{3} \; ) + {\epsilon _{0} c
\; \gamma \over \mu }  \; u^{0} \; (   u^{3} E_{2} - u^{2} E_{3})
\nonumber
\\
H_{2} =   {1  \over  \mu _{0} \mu }\;  B_{2}     +  {\gamma \over
\mu _{0} \mu } ( + u^{1} u^{2} B_{1}  -  u^{3} u^{3} \; B_{2}  -
u^{1} u^{1} \;  B_{2} +  u^{2} u^{3} B_{3} \; ) + {\epsilon _{0} c
\; \gamma \over \mu }  \; u^{0} \; (   u^{3} E_{2} -  u^{2} E_{3})
\nonumber
\\
H_{3} =   {1  \over  \mu _{0} \mu }\;  B_{3}     +  {\gamma \over
\mu _{0} \mu } ( + u^{3} u^{1} B_{1} +  u^{3} u^{2} B_{2} - u^{1}
u^{1} \;  B_{3}  - u^{2} u^{2} \;  B_{3}
 \; ) +
{\epsilon _{0} c  \; \gamma \over \mu }  \; u^{0} \; (   u^{2}
E_{1} -  u^{1} E_{2})
\nonumber
\end{eqnarray}

\noindent
Noting identities
\begin{eqnarray}
(-  u^{2} u^{2} \;  B_{1}  - u^{3} u^{3} \;  B_{1}  +  u^{1} u^{2}
B_{2} +  u^{1} u^{3} B_{3} \; )=
\nonumber
\\
= u^{2} ( u^{1}  B_{2} -  u^{2} \;  B_{1} ) - u^{3} ( u^{3}B_{1} -
u^{1}B_{3}) =
 [ {\bf u} \times ({\bf u} \times {\bf B} ) ]_{1}
\nonumber
\\
( + u^{1} u^{2} B_{1}  -  u^{3} u^{3} \;  B_{2}  - u^{1} u^{1} \;
B_{2} +  u^{2} u^{3} B_{3} \; ) =
\nonumber
\\
=   u^{3} ( u^{2}  B_{3}  -  u^{3} \;  B_{2} ) - u^{1} ( u^{1}
B_{2}  -  u^{2} \;  B_{1} )   =[ {\bf u} \times ({\bf u} \times
{\bf B} ) ]_{2}
\nonumber
\\
( + u^{3} u^{1} B_{1} +  u^{3} u^{2} B_{2} -  u^{1} u^{1} \; B_{3}
- u^{2} u^{2} \;  B_{3}
 \; ) =
 \nonumber
 \\
 = u^{1} (  u^{3}  B_{1}  - B_{1}U^{3})  )  - u^{2} (  u^{2} \;  B_{3} -
 u^{3} B_{2})  = [ {\bf u} \times ({\bf u} \times {\bf B} ) ]_{3}
 \nonumber
 \end{eqnarray}

\noindent
previous relations can be rewritten in a  vector form as follows:
\begin{eqnarray}
{\bf H} =   {1  \over  \mu _{0} \mu }\;  {\bf B}     +  {\gamma
\over  \mu _{0} \mu } { {\bf V} \times ({\bf V} \times {\bf B} ) )
\over 1 -V^{2}} + {\epsilon _{0} c  \; \gamma \over \mu  }  \; {
{\bf V} \times {\bf E} \over 1 -V^{2}} \label{130b}
\end{eqnarray}

Relations  (\ref{130a})-(\ref{130b})  provide us with
3-dimensional vector form of material equations  in media moving
with  velocity ${\bf V}$ . Firstly they were produced by H.
Minkowsky.

\section{Effective material equations generated by Riemannian \\
geometry of a space of constant positive  curvature}

A 3-dimensional space of constant positive  curvature, $S_{3}$,
has many applications in physical problems. The most simple
realization of this model is given by three-sphere in 4-space (the
space of the unitary group $SU(2)$ ):
\begin{eqnarray}
W_{4}^{2} + W_{1}^{2} +  W_{2}^{2} + W_{3}^{2}  = R^{2} \; ,
\qquad w_{1} = {W_{1} \over R} , \;\; \mbox{ and  so   on} \; ;
\label{A.1}
\end{eqnarray}

\noindent $R$ is a curvature radius.
These four coordinate are connected with quasi-spherical  ones be
relations:
\begin{eqnarray}
w_{4} = \cos \chi , \qquad w_{i} = \sin \chi \; n_{i} \; , \qquad
n_{i}= (\sin \theta \cos \phi,  \sin \theta \sin \phi, \sin \theta
) \; . \label{A.2}
\end{eqnarray}

The most used coordinates are conformally-flat ones:
\begin{eqnarray}
y^{i} = {2w_{i} \over 1 + w_{4}}  = 2 \tan \chi /2 \; n_{i} \; ,
\nonumber
\\
dl^{2} = R^{2} \; (1 + {y^{2} \over 4})^{-2} \; [dy_{1}^{2} +
dy_{2}^{2}  + dy_{3}^{2}  \; ] \label{A.3}
\end{eqnarray}

\noindent
and quasi-Cartesian ones :
\begin{eqnarray}
x^{i} = { y^{i} \over 1 - y^{2}/4} = \tan \chi \; n_{i} = {w_{i}
\over w_{4}} \; , \qquad \chi \in [0 , \pi /2 ] \; ; \label{A.4}
\end{eqnarray}

\noindent the later parameterize only elliptical model (the  space
of orthogonal group $S0(3)$ )
\begin{eqnarray}
dS^{2} = c^{2} dt^{2} - ({\delta_{jk} \over 1 + x^{2} } - { x^{j}
x^{k} \over (1 + x^{2})^{2} } ) \; dx^{j} dx^{j} \; , \nonumber
\\
g_{\alpha \beta} = \left | \begin{array}{cc}
1 & 0 \\
0 & g_{jk}
\end{array} \right | , \qquad g_{jk} = -
 ({\delta_{jk} \over 1 + x^{2} } - { x^{j} x^{k} \over (1 + x^{2})^{2} } ) \; ,
\nonumber
\\
g^{\alpha \beta} = \left | \begin{array}{cc}
1 & 0 \\
0 & g^{kl}
\end{array} \right | , \qquad g^{kl} = -(1 + x^{2})
 (\delta_{kl}  +  x^{k} x^{l}  ) \; .
\label{A5}
\end{eqnarray}

\noindent
Also calculating the determinant
\begin{eqnarray}
\mbox{det}\; (g_{\alpha \beta} ) = {1 \over  \mbox{det}\;
(g^{\alpha \beta} ) }, \qquad \mbox{det}\; (g^{\alpha \beta} )= -
(1+x^{2})^{3}
 \left | \begin{array}{ccc}
1 +x^{1}x^{1}  & x^{1}x^{2} & x^{1} x^{3} \\
x^{2} x^{1} & 1 +x^{2}x^{2} & x^{2} x^{3} \\
x^{3}x^{1} & x^{3}x^{2}  & 1 + x^{3} x^{3}
\end{array} \right | =
\nonumber
\\
- (1+x^{2})^{3} \; [ (1 +x^{1}x^{1} )(1 +x^{2}x^{2} )(1
+x^{3}x^{3} ) + 2x^{1}x^{1}  x^{2}x^{2}x^{3}x^{3} - \nonumber
\\
- (1 + x^{2}x^{2}) x^{1}x^{1} x^{3}x^{3} - (1 + x^{1}x^{1})
x^{2}x^{2} x^{3}x^{3} - (1 + x^{3}x^{3}) x^{1}x^{1} x^{2}x^{2} ] =
- (1+x^{2})^{3} \; , \label{A.6}
\end{eqnarray}

\noindent that is
\begin{eqnarray}
\sqrt{-\mbox{det}\; (g_{\alpha \beta} )} = {1 \over
(1+x^{2})^{3/2} } \; . \label{A.7}
\end{eqnarray}

For effective dielectric tensor  $\epsilon^{ik}(x)$   we have
\begin{eqnarray}
\epsilon^{ik}(x) = \sqrt{-g} \; g^{00} (x) g^{ik}(x) = - {1 \over
\sqrt{1+x^{2} }  }
 (\delta_{ik} + x^{i} x^{k}) =
 \nonumber
 \\
 =
- {1 \over \sqrt{1+x^{2}} } \left | \begin{array}{ccc}
1+x^{1}x^{1} & x^{1} x^{2} & x^{1} x^{3} \\
x^{1} x^{2} &  1+x^{2}x^{2} & x^{2} x^{3}\\
 x^{3} x^{1} & x^{3} x^{2} & 1+x^{3}x^{3}
\end{array}  \right | \; .
\label{A.8}
\end{eqnarray}

\noindent For effective magnetic  tensor   $\mu^{ik}(x)$ we have
\begin{eqnarray}
\mu^{ik}(x) =  \sqrt{1+x^{2}} \; \left | \begin{array}{ccc}
(1 + x^{2}x^{2} + x^{3}x^{3}) &  - x^{1}x^{2} & -x^{1}x^{3} \\
- x^{2}x^{1} & (1 + x^{3}x^{3} + x^{1}x^{1})  & -x^{2}x^{3} \\
-x^{3}x^{1} & -x^{3}x^{2}  & (1 + x^{1}x^{1} + x^{2}x^{2})
\end{array} \right |
\label{A.9}
\end{eqnarray}

It is easily verified by direct calculation that (taken with
minus ) dielectric tensor  $(- \epsilon ^{ik}(x)) $  and tensor
$\mu^{ik}(x)$ are inverse to each other

\begin{eqnarray}
-\epsilon^{ik}(x) \; \mu^{kl}(x) = \delta_{ik} \; . \label{A.10}
\end{eqnarray}

\noindent Let us write down the  effective material equations
explicitly
\begin{eqnarray}
D^{i} = \epsilon_{0} \epsilon^{ik} E_{k} \; , \qquad H^{i} = {1
\over \mu_{0} } \mu^{ik} B_{k},  \qquad B_{i} =  \mu_{0} M^{ik}
H^{k} \; ; \label{A.11}
\end{eqnarray}

\noindent at this two matrices coincide
\begin{eqnarray}
-\epsilon^{ik}(x) = M^{ik}(x) = {1 \over \sqrt{1+x^{2}} } \left |
\begin{array}{ccc}
1+x^{1}x^{1} & x^{1} x^{2} & x^{1} x^{3} \\
x^{1} x^{2} &  1+x^{2}x^{2} & x^{2} x^{3}\\
 x^{3} x^{1} & x^{3} x^{2} & 1+x^{3}x^{3} \; ,
 \end{array} \right |
 \label{A.12}
 \end{eqnarray}

\section{Effective material equations generated by Lobachevsky  \\ geometry of a space of constant negative   curvature}

A 3-dimensional space of constant  negative   curvature, $H_{3}$,
has many applications in physical problems. The most simple
realization of this model is given by three-sphere in 4-space (the
space of the unitary group $SU(1.1)$):
\begin{eqnarray}
W_{4}^{2} - W_{1}^{2} -  W_{2}^{2} - W_{3}^{2}  = R^{2} \; ,
\qquad w_{1} = {W_{1} \over R}, \;\; \mbox{ and  so   on} \; ;
\label{B.1}
\end{eqnarray}

\noindent $R$ is a curvature radius. These four coordinate are
connected with quasi-spherical  ones be relations:
\begin{eqnarray}
w_{4} = \cosh \chi , \qquad w_{i} = \sinh \chi \; n_{i} \; ,
\qquad n_{i}= (\sin \theta \cos \phi,  \sin \theta \sin \phi, \sin
\theta ) \; , \chi \in [ 0 , + \infty ) \; . \label{B.2}
\end{eqnarray}

The most used coordinates are conformally-flat ones:
\begin{eqnarray}
y^{i} = {2w_{i} \over 1 + w_{4}}  = 2 \tanh \chi /2 \; n_{i} \; ,
\nonumber
\\
dl^{2} = R^{2} \; (1 - {y^{2} \over 4})^{-2} \; [dy_{1}^{2} +
dy_{2}^{2}  + dy_{3}^{2}  \; ] \label{B.3}
\end{eqnarray}

\and quasi-Cartesian ones :
\begin{eqnarray}
x^{i} = { y^{i} \over 1 + y^{2}/4} = \tanh \chi \; n_{i} = {w_{i}
\over w_{4}} \; , \qquad \chi \in [0 , \pi /2 ] \; ; \label{B.4}
\end{eqnarray}

the later parameterize only elliptical model (the  space of
orthogonal group $S0(3)$ )
\begin{eqnarray}
dS^{2} = c^{2} dt^{2} - ({\delta_{jk} \over 1 - x^{2} } + { x^{j}
x^{k} \over (1 - x^{2})^{2} } ) \; dx^{j} dx^{j} \; , \nonumber
\\
g_{\alpha \beta} = \left | \begin{array}{cc}
1 & 0 \\
0 & g_{jk}
\end{array} \right | , \qquad g_{jk} = -
 ({\delta_{jk} \over 1 - x^{2} } + { x^{j} x^{k} \over (1 - x^{2})^{2} } ) \; ,
\nonumber
\\
g^{\alpha \beta} = \left | \begin{array}{cc}
1 & 0 \\
0 & g^{kl}
\end{array} \right | , \qquad g^{kl} = -(1 - x^{2})
 (\delta_{kl}  -  x^{k} x^{l}  ) \; .
\label{B.5}
\end{eqnarray}

\noindent
Also calculating the determinant
\begin{eqnarray}
\mbox{det}\; (g_{\alpha \beta} ) = {1 \over  \mbox{det}\;
(g^{\alpha \beta} ) }, \qquad \mbox{det}\; (g^{\alpha \beta} )=-
(1-x^{2})^{3} \; , \label{B.6}
\end{eqnarray}

\noindent that is
\begin{eqnarray}
\sqrt{-\mbox{det}\; (g_{\alpha \beta} )} = {1 \over (1 -
x^{2})^{3/2} } \; . \label{B.7}
\end{eqnarray}

For effective dielectric tensor  $\epsilon^{ik}(x)$   we have
\begin{eqnarray}
\epsilon^{ik}(x) = \sqrt{-g} \; g^{00} (x) g^{ik}(x) = - {1 \over
\sqrt{1-x^{2} }  }
 (\delta_{ik} - x^{i} x^{k} =
 \nonumber
 \\
 =
- {1 \over \sqrt{1-x^{2}} } \left | \begin{array}{ccc}
1-x^{1}x^{1} & -x^{1} x^{2} & -x^{1} x^{3} \\
-x^{1} x^{2} &  1-x^{2}x^{2} & -x^{2} x^{3}\\
 -x^{3} x^{1} & -x^{3} x^{2} & 1-x^{3}x^{3}
\end{array}  \right | \; .
\label{B.8}
\end{eqnarray}

\noindent For effective magnetic  tensor   $\mu^{ik}(x)$ we have
\begin{eqnarray}
\mu^{ik}(x) = \sqrt{-g} \left | \begin{array}{ccc}
(g^{22}g^{33} - g^{23}g^{32}) & (g^{31}g^{23} - g^{21}g^{33} ) &  (g^{21}g^{32} - g^{22}g^{31} )\\
(g^{32}g^{13} - g^{33}g^{12} ) & (g^{33}g^{11} - g^{31}g^{13} ) & (g^{31}g^{12} - g^{32}g^{11} )\\
(g^{12}g^{23} - g^{13}g^{22} )  & (g^{13}g^{21} - g^{11}g^{23} ) &
(g^{11}g^{22} - g^{12}g^{21} )
\end{array} \right |=
\nonumber
\\
= \sqrt{1-x^{2}} \; \left | \begin{array}{ccc}
(1 - x^{2}x^{2} - x^{3}x^{3}) &   x^{1}x^{2} & x^{1}x^{3} \\
 x^{2}x^{1} & (1 - x^{3}x^{3} - x^{1}x^{1})  & x^{2}x^{3} \\
x^{3}x^{1} & x^{3}x^{2}  & (1 - x^{1}x^{1} - x^{2}x^{2})
\end{array} \right |
\label{B.9}
\end{eqnarray}

It is easily verified by direct calculation that (taken with
minus ) dielectric tensor  $(- \epsilon ^{ik}(x)) $  and tensor
$\mu^{ik}(x)$ are inverse to each other
\begin{eqnarray}
-\epsilon^{ik}(x) \; \mu^{kl}(x) = \delta_{ik} \; . \label{B.10}
\end{eqnarray}

\noindent Let us write down the  effective material equations
explicitly
\begin{eqnarray}
D^{i} = \epsilon_{0} \epsilon^{ik} E_{k} \; , \qquad H^{i} = {1
\over \mu_{0} } \mu^{ik} B_{k},  \qquad B_{i} =  \mu_{0} M^{ik}
H^{k} \; ; \label{B.11}
\end{eqnarray}

\noindent at this two matrices coincide
\begin{eqnarray}
-\epsilon^{ik}(x) = M^{ik}(x) = {1 \over \sqrt{1-x^{2}} } \left |
\begin{array}{ccc}
1-x^{1}x^{1} & -x^{1} x^{2} & -x^{1} x^{3} \\
-x^{1} x^{2} &  1-x^{2}x^{2} & -x^{2} x^{3}\\
 -x^{3} x^{1} & -x^{3} x^{2} & 1-x^{3}x^{3} \; ,
 \end{array} \right | \; .
 \label{B.12}
 \end{eqnarray}

\section{Geometry effect on material equations in media
}

Above, we have started with Maxwell equations in vacuum:
\begin{eqnarray}
\partial_{a} F_{b c} + \partial_{b} F_{ca} +
\partial_{c} F_{ab} =0 \; ,
\nonumber
\\
  \partial_{b}
H^{ ba} =   j^{a} \; ,  \qquad H_{a b}  = \epsilon_{0} \; F_{ab}
\label{15.1}
\end{eqnarray}

\noindent and changed them to generally covariant in Riemannian
space-time
\begin{eqnarray}
\partial_{\alpha} F_{\beta \gamma} + \partial_{\beta} F_{\gamma \alpha }+
\partial_{\gamma} F_{\alpha \beta } =0 \; ,
\qquad
 {1 \over \sqrt{-g}}\;  {\partial_{\beta} \sqrt{-g} \;
H^{ \beta \alpha} =   j^\alpha} \; , \label{15.2a}
\end{eqnarray}

\noindent At this, vacuum material equations
\begin{eqnarray}
H_{\alpha \beta }  = \epsilon_{0} \; F_{\alpha \beta} ,
\label{15.2b}
\end{eqnarray}

\noindent due to presence of metrical tensor  $g^{\rho \alpha }
(x)$ gave us the modified (effective) material equations
\begin{eqnarray}
H^{\rho  \sigma }(x)  = \sqrt{-g}\; g^{\rho \alpha } (x) g^{\sigma
\beta}(x) \; \epsilon_{0} \; F_{\alpha \beta}(x) \;  .
\label{15.2c}
\end{eqnarray}

\vspace{5mm}

{\em As a first  generalization let us start with Maxwell
equations in a uniform media :}

\begin{eqnarray}
\partial_{a} F_{b c} + \partial_{b} F_{ca} +
\partial_{c} F_{ab} =0 \; ,
\qquad   \partial_{b} H^{ ba} =   j^{a} \; , \label{15.3a}
\\
\qquad H_{mn }  = \epsilon_{0}    \; \eta_{m}^{\;\; a} \;
\eta_{n}^{\;\; b} \; \; F_{ab}, \qquad \eta_{m}^{\;\; a} = \sqrt{\epsilon}\left |
\begin{array}{cccc}
k^{-1} & 0 & 0 & 0 \\
0 & k & 0 & 0 \\
0 & 0 & k & 0 \\
0 & 0 & 0 & k
\end{array} \right |\; , \;\; k^{2} = {1 \over \epsilon \mu } \;.
\label{15.3b}
\end{eqnarray}

Extension of these to Riemannian space-time looks as
\begin{eqnarray}
\partial_{\alpha} F_{\beta \gamma} + \partial_{\beta} F_{\gamma \alpha }+
\partial_{\gamma} F_{\alpha \beta } =0 \; ,
\qquad
 {1 \over \sqrt{-g}}\;  {\partial_{\beta} \sqrt{-g} \;
H^{ \beta \alpha} =   j^\alpha} \; , \label{15.4a}
\end{eqnarray}

\noindent At this, material equations for the uniform media
\begin{eqnarray}
\qquad H_{\alpha \beta }(x)  = \epsilon_{0}    \;
\eta_{\alpha}^{\;\; a} \; \eta_{\beta}^{\;\; b} \;
 F_{ab} (x) \; ,
\qquad \eta_{\alpha}^{\;\; a} = \sqrt{\epsilon} \left | \begin{array}{cccc}
k^{-1} & 0 & 0 & 0 \\
0 & k & 0 & 0 \\
0 & 0 & k & 0 \\
0 & 0 & 0 & k
\end{array} \right |\;.
\label{15.4b}
\end{eqnarray}

\noindent
  will  take the form
\begin{eqnarray}
H^{\rho  \sigma }(x) (x)  = \sqrt{-g}\; g^{\rho \alpha } (x)
g^{\sigma \beta}(x) \;  H_{\alpha \beta}(x) =  \nonumber
\\
=
 \sqrt{-g}\; g^{\rho \alpha } (x)
g^{\sigma \beta}(x) \; \epsilon_{0}  \;
   \; \eta_{\alpha }^{\;\; a} \; \eta_{\beta}^{\;\; b} \; \; F_{ab}(x) \; .
\label{15.4c}
\end{eqnarray}

\noindent With the notation
\begin{eqnarray}
\hat{F}_{\alpha \beta}(x)   =  \eta_{\alpha }^{\;\; a}
\; \eta_{\beta}^{\;\; b} \; \; F_{ab}(x)\; ; \nonumber
\end{eqnarray}

\noindent they are written as follows
\begin{eqnarray}
H^{\rho  \sigma }(x) (x)  =
 \sqrt{-g}\; g^{\rho \alpha } (x)
g^{\sigma \beta}(x) \; \epsilon_{0}  \; \hat{F}_{\alpha \beta}(x)
\; . \; . \label{15.4d}
\end{eqnarray}

\noindent where explicit form of $\hat{F}_{\alpha \beta}(x)$ is
\begin{eqnarray}
\hat{F}_{\alpha \beta}(x) = \left | \begin{array}{cc}
0                & \epsilon F_{0i} \\
\epsilon F_{i0}  &  \mu ^{-1} F_{ik}
\end{array} \right | .
\label{15.4e}
\end{eqnarray}

One should not make any additional calvulation, instead it
suffices the make  one formal change $F_{\alpha \beta}(x)
\Longrightarrow \hat{F}_{\alpha \beta}(x) $, and now material
equations are
\begin{eqnarray}
D^{i}=  \epsilon_{0} \epsilon \; \epsilon^{ik}(x) \; E_{k}+
\epsilon_{0} \epsilon c\; \alpha^{ik}(x) \; B_{k} \; , \nonumber
\\
H^{i}  = \epsilon_{0} \epsilon  c \; \beta^{ik}(x) \; E_{k}+ {1
\over \mu _{0} \mu}  \; \mu^{ik}(x) \; B_{k} \; . \label{15.5a}
\end{eqnarray}

These relations provide us with material equations for uniform
media modified by Riemannian geometry of background space-time.

It is easily to male one other extension: let us start with
anisotropic media in Minkowsky space
\begin{eqnarray}
D_{i} = \epsilon_{0} \; \epsilon^{(0)}_{kl} E_{l} \; , \qquad
H_{i} ={1 \over \mu_{0} }\; \mu^{(0)}_{kl} B_{k} \; .
\label{15.6a}
\end{eqnarray}

\noindent they will be modified into
\begin{eqnarray}
D^{i}=  \epsilon_{0} \; [ \epsilon^{ik}(x) \;  \epsilon^{(0)}_{kl}
] \;\; E_{l}+ \epsilon_{0}  c\;  [ \alpha^{ik}(x) \;
\mu^{(0)}_{kl} ] \;\;  B_{l}  \; , \nonumber
\\
H^{i}  = \epsilon_{0}  c \; [ \beta^{ik}(x) (\epsilon^{(0)}_{kl}]
\;\;  E_{l} +
 {1 \over \mu _{0} }  \;[  \mu^{ik}(x)   (\mu^{(0)}_{kl} ] \;\; B_{l}  \; .
\label{15.6b}
\end{eqnarray}

And now, final extension: let start with arbitrary (linear) media
when material equations are determined by 4-rank tensor
\begin{eqnarray}
H_{\alpha \beta} (x) = \epsilon_{0} \; \Delta_{\alpha
\beta}^{\;\;\;\;\;ab} \; F_{ab}(x) \label{15.7a}
\end{eqnarray}

\noindent
from which Rimannian geometry will generate the following ones
\begin{eqnarray}
H^{\rho  \sigma }(x) (x)  =
 \sqrt{-g}\; g^{\rho \alpha } (x)
g^{\sigma \beta}(x) \; \epsilon_{0} \; \Delta_{\alpha
\beta}^{\;\;\;\;\;ab} \; F_{ab}(x) \; . \label{15.8a}
\end{eqnarray}

\noindent or in 3-dimensional form
\begin{eqnarray}
D^{i}=  \epsilon_{0} \;  \epsilon^{ik}(x) \; \left
[\epsilon^{(0)}_{kl} \; E_{l} + c \alpha^{(0)}_{kl}\; B_{l}
\right ] + \epsilon_{0} \; \alpha^{ik}(x) \;  \left [
\beta^{(0)}_{kl} \; E_{l}  + \mu^{(0)}_{kl} c\; B_{l}  \right ]
\; , \nonumber
\\
H^{i}  = \epsilon_{0}  c  \;   \beta^{ik}(x)   \; \left
[\epsilon^{(0)}_{kl} \; E_{l} + c \alpha^{(0)}_{kl}\; B_{l} \right
]   +
 \epsilon_{0}  c   \; \mu^{ik}(x)\; \left  [  \beta^{(0)}_{kl} \; E_{l}
  + \mu^{(0)}_{kl} c \; B_{l} \right  ]  \; ;
\label{15.8b}
\end{eqnarray}

\noindent these may be rewritten differently
\begin{eqnarray}
D^{i}=  \epsilon_{0} \; \left [ \epsilon^{ik}(x) \;
\epsilon^{(0)}_{kl}  + \alpha^{ik}(x)  \beta^{(0)}_{kl}    \right
]   \; E_{l} + \epsilon_{0}c \; \left [ \epsilon^{ik}(x)
\alpha^{(0)}_{kl} + \alpha^{ik}(x) \mu^{(0)}_{kl}  \right ] \;
B_{l} \; ,
\nonumber
\\
H^{i}  = \epsilon_{0}  c  \;     \; \left [  \beta^{ik}(x)
\epsilon^{(0)}_{kl}  + \mu^{ik}(x) \beta^{(0)}_{kl}  \right ] \;
E_{l} + {1 \over \mu_{0}} \left [ \beta^{ik}(x) \alpha^{(0)}_{kl}
+ \mu^{ik}(x) \mu^{(0)}_{kl}  \right ] \; B_{l} \; , \nonumber
\end{eqnarray}

\noindent  or in   matrix form (with no indices)
\begin{eqnarray}
{\bf D}  =  \epsilon_{0} \; \left [ \epsilon (x) \;
\epsilon^{(0)}  + \alpha (x)  \beta^{(0)}    \right  ]   \;{\bf  E
} + \epsilon_{0}c \; \left [ \epsilon(x) \alpha^{(0)}  + \alpha(x)
\mu^{(0)}  \right ] \; {\bf B }\; , \nonumber
\\
{\bf H } = \epsilon_{0}  c  \;     \; \left [  \beta (x)
\epsilon^{(0)}  + \mu (x) \beta^{(0)}  \right ] \; {\bf E}   + {1
\over \mu_{0}} \left [ \beta(x) \alpha^{(0)} + \mu(x) \mu^{(0)}
\right ] {\bf B} \; . \label{15.8c}
\end{eqnarray}

These  formulas can be read symbolically:
\begin{eqnarray}
\epsilon^{0} \Longrightarrow \hat{\epsilon} =  \epsilon (x) \;
\epsilon^{(0)}  + \alpha (x)  \beta^{(0)}    , \qquad \alpha^{0}
\Longrightarrow \hat{\alpha} =
 \epsilon(x) \alpha^{(0)}  + \alpha(x) \mu^{(0)} \; ,
\nonumber
\\
\beta ^{0}\Longrightarrow  \hat{\beta} =
  \beta (x)  \epsilon^{(0)}  +
\mu (x) \beta^{(0)} \; , \qquad  \qquad  \mu ^{0} \Longrightarrow
\hat{\mu}=
 \beta \alpha^{(0)} + \mu(x) \mu^{(0)} \; .
\label{15.8d}
\end{eqnarray}

\noindent For instance, if starting material equations have only
diagonal blocks, that is  $\alpha^{0}=0, \beta^{0}=0$, last
relations become simpler:
\begin{eqnarray}
\epsilon^{0}  \Longrightarrow \hat{\epsilon} =  \epsilon (x) \;
\epsilon^{(0)}     ,
 \qquad \qquad \alpha^{0}=0  \Longrightarrow
\hat{\alpha} =   + \alpha(x) \mu^{(0)} \; , \nonumber
\\
\beta^{0}=0 \Longrightarrow  \hat{\beta} =
  \beta (x)  \epsilon^{(0)}   \;
, \qquad  \mu^{0}   \Longrightarrow  \hat{\mu}=
  \mu(x) \mu^{(0)} \; .
\label{15.8e}
\end{eqnarray}

Four (material) tensor in the above formulas are defined by

\end{document}